# How to improve the outcome of performance evaluations in terms of percentiles for citation frequencies of my papers

**Michael Schreiber**

*Institute of Physics, Chemnitz University of Technology, 09107 Chemnitz, Germany.*
*E-mail: schreiber@physik.tu-chemnitz.de*

Using empirical data I demonstrate that the result of performance evaluations by percentiles can be drastically influenced by the proper choice of the journal in which a manuscript is published.

## 1. Introduction

In order to evaluate the impact of publications in terms of citation counts recently percentile-based bibliometric indicators have gained more and more attention (Bornmann, Leydesdorff, & Mutz, 2013). They are based on the idea of determining the position of a publication within a citation distribution of its field and its publication year (Leydesdorff & Bornmann, 2011). This position is usually expressed in terms of a quantile of the citation distribution. Commonly, percentiles are used for that purpose. These can then be utilized to attribute the evaluated papers to certain percentile rank classes. A simple and frequently used indicator is the proportion of publications which belong to the top 10% most frequently cited publications (Tijssen, Visser & Van Leeuwen, 2002, Bornmann, De Moya Anegón & Leydesdorff, 2012) in a given field for a given publication year. This would mean two percentile rank classes, namely the top 10% and the bottom 90%. Of course, different thresholds might be useful for different evaluation purposes (Lewison, Thornicroft, Szmukler & Tansella, 2007). Also more complicated indicators combining several thresholds and several percentile rank classes have been proposed (Leydesdorff, Bornmann, Mutz & Opthof 2011, Leydesdorff & Bornmann, 2012).

The much appraised advantage of percentile-based indicators is that they allow a straightforward comparison of the evaluation results for publications in different fields and a suitable comparison of old and new publications, because the citation frequency of each publication of an author (or an institute, or a country, …) is compared with the complete citation distribution of publications in the corresponding field and publication year. It is well known that these reference sets can differ strongly between different fields. Therefore it appears fair to measure the impact of a publication in this way, i.e. in relation to its field. By the same token this means, however, that one can influence the position of one's publication in its citation distribution by choosing the journal and thereby the field and thus the reference set in a more or less clever way. It is the purpose of the present paper to demonstrate this effect for an empirical example, namely five of my own publications for which a choice of a different journal would have drastically improved the position of the paper in the respective citation distribution in each of the five cases.

I have recently already found a strong influence of the journal or rather the subject category on the performance evaluation results (Schreiber, 2014) in terms of the new citation-rank indicator P100 which is also based on percentiles, but utilizes the distribution of unique citation values (Bornmann, Leydesdorff, & Wang, 2013). Plotting the P100 indicator values of 244 papers I was surprised by some outliers that appeared below the main body of data points in the plot. A closer inspection revealed that two outliers belong to the subject category "Chemistry Multidisciplinary", two other outliers were attributed to the subject category "Multidisciplinary Sciences", while most papers in the main stream of data were from "Information Science Library Science" (hereafter abbreviated as InfScLibSc).

## 2. Performance evaluation for five articles in multidisciplinary physics

I am a physicist, but I became interested in bibliometric studies after another physicist, Jorge Hirsch proposed the h-index as a simple indicator for measuring the impact of the publications of a researcher. I performed several case studies, sometimes analyzing small fictitious datasets, but often using empirical datasets comprising citation data of more or less famous physicists or of a physics journal. Thus bibliometrics became my research hobby horse. However, being a physicist I looked at my habitual turf when I had to choose a journal to which to submit the manuscripts. Thus my first three publications on variants of the h-index were published in physics journals (Schreiber, 2007a, 2007b, 2008a). This choice was indolent and easy, but it seemed appropriate, because I analyzed datasets from physics. Later I published two more papers on the h-index in physics journals (Schreiber, 2009, 2010), but more than twenty further papers in journals which are attributed to information science and library science.

For the subsequent investigation I have downloaded and analyzed citation distributions from Thomson Reuters' Web of Science on 1 and 2 July 2014 for the subject category "Physics Multidisciplinary" (hereafter abbreviated as PhyMu) to which the mentioned five publications are solely attributed. Table 1 shows that these papers have received a gratifying number of citations.

I have sorted the citation distributions by *decreasing* number of citations. Scaled by the total number of articles in the reference set the position in the distribution determines a percentage value for a publication, which is then used to assign the paper to a percentile. In this way the lowest percentile will be attributed to the most cited paper and the highest percentile to the least cited (usually uncited) papers. Consequently, the $10^{th}$ percentile determines for example the threshold which distinguishes the top 10% most frequently cited publications from the bottom 90%.

Percentiles are often determined after sorting the citation distribution by *increasing* number of citations. Here I have decided to do it the other way round, because then the percentage $x$ directly reflects the top $x$% most frequently cited papers. The same sequence is used for the percentage scale of the InCites data in the Web of Science (Bornmann, Leydesdorff & Wang, 2013). The resulting percentiles are sometimes distinguished by calling them *inverted percentiles*.

There are often several papers in a reference set which have received the same number of citations. It is not obvious how to sort these tied items. In Table 1 the respective range of ranks and the corresponding range of percentage values is given. Some time ago there has been an intense discussion in the literature about the best way how to assign tied publications to percentiles and thus to percentile rank classes (Waltman & Schreiber, 2013; and references therein). One can summarize that debate in such a way that the percentage range in the table can be interpreted as an uncertainty interval which covers the different proposals (Leydesdorff, 2012). The possibilities include to utilize the top (Leydesdorff & Bornmann, 2011, Leydesdorff, Bornmann, Mutz & Opthof 2011) or the bottom (Rousseau, 2012) of the range, or the average (Pudovkin & Garfield, 2009) of the percentages to assign a percentile and thus a percentile rank class. If one (or even more than one) threshold between percentile rank classes falls into the percentage range, the different possibilities lead to different results (Schreiber, 2013a, 2013b, 2013c). In my view in such a case a fractional attribution to different percentile rank classes (Schreiber, 2012, Waltman & Schreiber, 2013) is the best solution. The rather large number of tied papers in each of the five cases in Table 1 shows that this is not a purely academic problem; for example it is unclear whether the 2009 paper falls into the top quartile (top 25%) or not. For smaller citation frequencies the number of tied publications becomes much larger and thus also the uncertainty intervals for the percentage grow strongly. But I do not want to enter that debate (Waltman & Schreiber, 2013, Schreiber, 2013b) here and therefore I restrain myself from using percentiles and percentile rank classes in the further discussion, but I rather compare percentage ranges or the median of the percentages attributed to the tied publications, which is given in Table 1, too. It can easily be determined as the arithmetic average of the top and the bottom of the percentage range. It equals also the average over all percentage values in that range.

### 3. Comparison with three other subject categories

Eight of my subsequent manuscripts in bibliometrics have been published in the Journal of Informetrics which is attributed in the Web of Science to the subject category InfScLibSc. I have also downloaded the respective citation distributions from the Web of Science and determined the ranges of ranks for the same numbers of citations and the same publication years as in Table 1, for this subject category, see Table 2. The total number of articles in these five reference sets is smaller than in Table 1 by more than one order of magnitude (on average 13.5 times). But this is not so important. More important is that the highest and lowest ranks for the tied publications are smaller by about one and a half order of magnitude (on average 31 times). Consequently the percentage range shifts to much smaller values. Comparing the median percentages in Table 2 with the respective values in Table 1 one can see that the same number of citations in the PhyMu category means a higher percentage (on average 2.3 times higher) than in InfScLibSc.

In conclusion, if the five papers from Table 1 had been published in the Journal of Informetrics, the evaluation would lead to much lower percentage values and thus to lower (inverted) percentiles, which means that a better result would come out in a performance evaluation based on such percentiles. Of course, this conclusion depends on the hypothesis that the publications would have received the same numbers of citations irrespective of the chosen journal, i.e. the absolute impact is assumed to be independent of the subject category. But the relative impact in comparison with the corresponding reference set does strongly depend on the respective subject category.

A similar effect can be detected for the Journal of the American Society of Information Science and Technology. In the mean time I have published nine manuscripts in this journal which is attributed in the Web of Science also to the subject category InfScLibSc as well as to the subject category "Computer Science Information Systems" (ComScInfSy). Therefore I have analyzed the citation distributions and determined the ranges of ranks for the tied publications with the same number of citations as in Table 1 and the same publication years for articles in the latter subject category. The results are given in Table 3 and the numbers are between the corresponding numbers in Tables 1 and 2, but closer to the latter.

In conclusion, if I had chosen a journal from this subject category for the five publications, and if the papers would have received the same number of citations, again the performance evaluation would have led to a much better result (on average two times better). For the particular journal which is attributed to two subject categories one would usually average the percentiles, which means that the averages of the percentages in Tables 2 and 3 would become relevant yielding better values than Table 3 alone. Therefore to be specific, a publication in that journal instead of the physics journals would have made my performance look better (on average by a factor of 2.15).

Usually the observed effects can be expected to be stronger between different disciplines like physics and scientometrics rather than between neighboring subject categories in the same field. This is exemplified by comparing Table 1 with Table 2 or Table 3 for the first case of different disciplines and comparing Tables 2 and 3 for the second case of neighboring subject categories in the same field.

However, this observation should not be generalized. As a counterexample I have also investigated the subject category "Computer Science Interdisciplinary Applications" (ComScIntAp). The results are presented in Table 4. The median percentage values are (on average 1.4 times) better than for PhyMu, but (on average 1.4 times) worse than in the subject category ComScInfSy and (on average 1.6 times) worse than in InfScLibSc. Here the differences between Table 1 and Table 4, i.e. between different disciplines, are as large as the differences between Tables 3 and 4, i.e. between neighboring subject categories in the same field. I note that the well known journal Scientometrics which I have also utilized for my publications in this field, is categorized in the Web of Science as belonging to ComScIntAp as well as to InfScLibSc. In conclusion, choosing this journal instead of the physics journals for the five publications in Table 1, the performance would have also looked better (on average by a factor of 1.8), but not as good as in Journal of Informetrics.

To be more specific, my first publication on the g-index was published in Scientometrics (Schreiber, 2008b) and has been cited 32 times. In Table 5 I compare the respective ranks and percentages for the 4 subject categories discussed above. Similar observations can be made: The same number of citations leads to large deviations in the percentages for the different categories. The sequence of the performance evaluation results is the same as above. Consequently, assuming again the same number of citations, in one of the PhyMu journals the paper would have fared (1.88 times) worse than in Scientometrics, but in the Journal of Informetrics it would be evaluated (1.25 times) better and also (1.20 times) better in the Journal of the American Society for Information Science and Technology.

## 4. Visualization

In order to visualize the observed behavior, I present the citation distributions for articles published in 2007 in the 4 discussed subject categories in Figure 1. The differences in the number of articles but also in the number of citations are obvious. For example the short *vertical* bar at the beginning of the PhyMu curve for the citation frequency 1 ranges from 33058 papers with more than 0 citations to 29340 papers with more than 1 citation. In contrast, the InfScLibSc curve drops from only 2101 cited papers to 1786 papers with more than 1 citation. The following horizontal bars are so long, because they reflect the integer step to the citation frequency 2 and this step appears so long due to the logarithmic scale of the horizontal axis. Relevant for the discussion is the next vertical step at citation frequency 2 which reflects the tied papers with 2 citations. For the PhyMu curve it ranges from the above number of 29340 papers with more than 1 citation to the number of 26257 papers with more than 2 citations. Here, the InfScLibSc curve drops from 1786 to 1550 papers with more than 2 citations. Due to logarithmic scale of the vertical axis these vertical bars are relatively short.

At the other end of the curve the logarithmized values lead to long vertical bars although the numbers decrease only by one paper for these high citation frequencies. On the other hand, in this regime long horizontal bars are conspicuous. These reflect the many high citation frequencies which do not occur in the actual distribution. Therefore, the number of more frequently cited articles does not change at these citation frequencies. The end of the PhyMu curve at citation frequency 9966 denotes the most cited paper. There are 0 papers with more than 9966 citations so that the curve drops here to minus infinity due to the logarithmic scale of the vertical axis. The InfSciLibSc curve ends at only 315 citations to the most cited paper.

Due to the logarithmic scale of the vertical axis, the normalization with respect to the total number of articles means only a downward shift of all curves so that all of them begin at 100%. The result is shown in Figure 2. In fact, the displayed curves do not begin at 100%, because the respective citation frequency 0 would correspond to minus infinity on the logarithmic scale of the horizontal axis. Interestingly, the sequence of the curves in Figure 2 remains the same as in Figure 1, except for ComScIntAp which lies above the PhyMu curve for small citation frequencies and close to it for high citation frequencies.

In Figure 3 I have visualized the data which are most relevant for the above analysis in a different way. Here the numbers of citations are presented versus the percentages, on linear scales. This is the way in which citation distributions are usually displayed, except for the here performed normalization of the horizontal axis. The citation frequencies 26 and 45 which have been utilized in Tables 1 – 4 above are indicated in the figure. The differences between the percentages in the different reference sets for the specific citation frequencies are eye-catching.

## 5. Discussion

It is a well-known fact that citation distributions can differ strongly between different fields, as shown in Figure 1. Consequently, a meaningful comparison across the fields requires a field-dependent normalization as performed in Figure 2. On first sight, it appears that the large differences between the curves in Figure 1 have been more or less compensated as expected. However, due to the different shapes of the curves a closer inspection of Figure 2 reveals that the remaining deviations are quite relevant for the evaluation of the impact of a publication in terms of its position in the citation distribution of a reference set. Of course this does not mean that percentiles cannot be compared across fields, but it means that it can make a large difference whether a paper is attributed to one reference set or another.

One possibility of selecting a reference set for an article is to utilize all papers published in the same journal in the same year. However, this rewards publications in journals of little reputation (Bornmann, Marx, & Barth, 2013). It also leads to relatively small reference sets for most journals. A better way is based on the classification in terms of research areas, e.g. the so-called subject categories in the Web of Science. These are constructed on the basis of journals, which are attributed to one or to several subject categories. Here the assumption is that journals contain a set of papers which is coherent in topics (Bornmann, Marx, & Barth, 2013). That is, however, problematic for multidisciplinary journals. The way out of this difficulty is a separate category for these journals. Nevertheless, field normalization remains questionable when it is based on journal classification schemes, even if journal sets instead of individual journals are used for the construction of the reference datasets.

There has been a long discussion of this problem in bibliometrics and various alternative classification systems and techniques for their construction habe been suggested and tested (Waltman & van Eck, 2012). The preferred method of choice is the categorization on the level of individual publications. This can be based on keywords from a predefined list, like the Medical Subject Headings utilized by Leydesdorff and Opthof (2013) or the Chemical Abstract sections discussed by Bornmann, Marx, and Barth (2013) or the Physics and Astronomy Classification System (PACS). An interesting new methodology has recently been proposed by Waltman and van Eck (2012); it is based on citation relations and leads to clustering of publications into research areas which correspond only partially to traditional disciplines. This approach can be refined further by exploiting other relations besides the citations between the papers, e.g. shared words in titles and abstracts. For the field-specific normalization of citation distributions, there seems at present to be no generally accepted classification scheme on a paper-by-paper basis. Moreover, there are also practical limitations, if such a classification system has to be constructed for a very large number of papers (Waltman & van Eck, 2012). Another practical problem is the necessity to combine the classification scheme with the citation records in the Web of Science or other citation data bases (Leydesdorff & Opthof, 2013).

The analysis in the previous section could therefore not be based on a publication-based classification scheme, but was performed in terms of the subject categories from the Web of Science. These are journal based. But that does not mean that the field-specific normalization of the citation distributions is based on an individual journal. Rather several journals contribute to one subject category and on the other hand some journals are attributed to more than one subject category. This makes a straightforward conclusion about the influence of the journal on the normalized citation scores impossible. Nevertheless, there can be a large influence.

The empirical data in the previous section have shown how important it can be which journal one chooses for the publication of a manuscript. The proper choice can drastically change the resulting position in the corresponding citation distribution and thus drastically alter the outcome of performance evaluations in terms of percentiles. A good publication strategy can improve the measured performance.

Of course adjusting the publication strategy to performance indicators is nothing new. Already decades ago people slandered about researchers who subdivided their research results into "least publishable units" (LPUs) in order to enhance the number of their publications, because in those days it became fashionable to count the number of publications of a researcher. More recently, the popularity of the h-index as an indicator, however questionable, of scientific achievements suggests another publication strategy: Selfcitations to papers which are just below one's value of the h-index are an easy way of increasing one's index value.

Concerning the proper choice of the journal for the best values of percentile-based indicators, I have already mentioned that I have assumed for the above investigation that the number of citations would not depend on the chosen journal. This is of course a strong condition. However, for the present examples it is unlikely that the considered articles would have received less citations in bibliometric journals than in the physics journals. One could even say that the articles were hidden in the physics journals and that they would possibly have attracted more attention and maybe more citations if they had been published in bibliometric or scientometric journals and thus received more attention and more citations from researchers in these fields. But this supposition should not be generalized: If one selects a journal which is attributed to a particularly attractive subject category because the citation frequencies of articles in that subject category are comparatively low, this could mean that (nearly) nobody reads the publication, it gets less attention and receives less citations. Thus this would be a counterproductive strategy. The effect could be even worse if the selected journal belongs to a subject category which is not in the range of vision of most researchers in the field. The last argument is valid in spite of possible computerized searches, because many people do not extend those searches to fringe journals.

Another caveat is that for the proper choice of the journal one would need a prediction how the citation distribution of the journal or rather the corresponding subject category will develop in the future.

**Fig. 1.** Citation distributions for articles published in 2007 in one of the 4 subject categories which are given in the inset. Note the doubly logarithmic scale. The vertical pieces of the curves reflect tied papers, i.e. articles with the same number of citations. Thus the curves begin on the top left with the number of cited articles and end on the bottom right with the highest number of citations.

**Fig. 2.** Same as Fig. 1, but scaled by the total number of articles in each reference set. The data for the singly cited articles have been slightly shifted to the right, so that the values can be better distinguished.

**Fig. 3.** Citation frequencies versus percentages on linear scales. Horizontal pieces of the curves reflect tied publications with the same number of citations. The two horizontal lines indicate the citation frequencies which have been discussed and utilized for Tables 1 – 4.

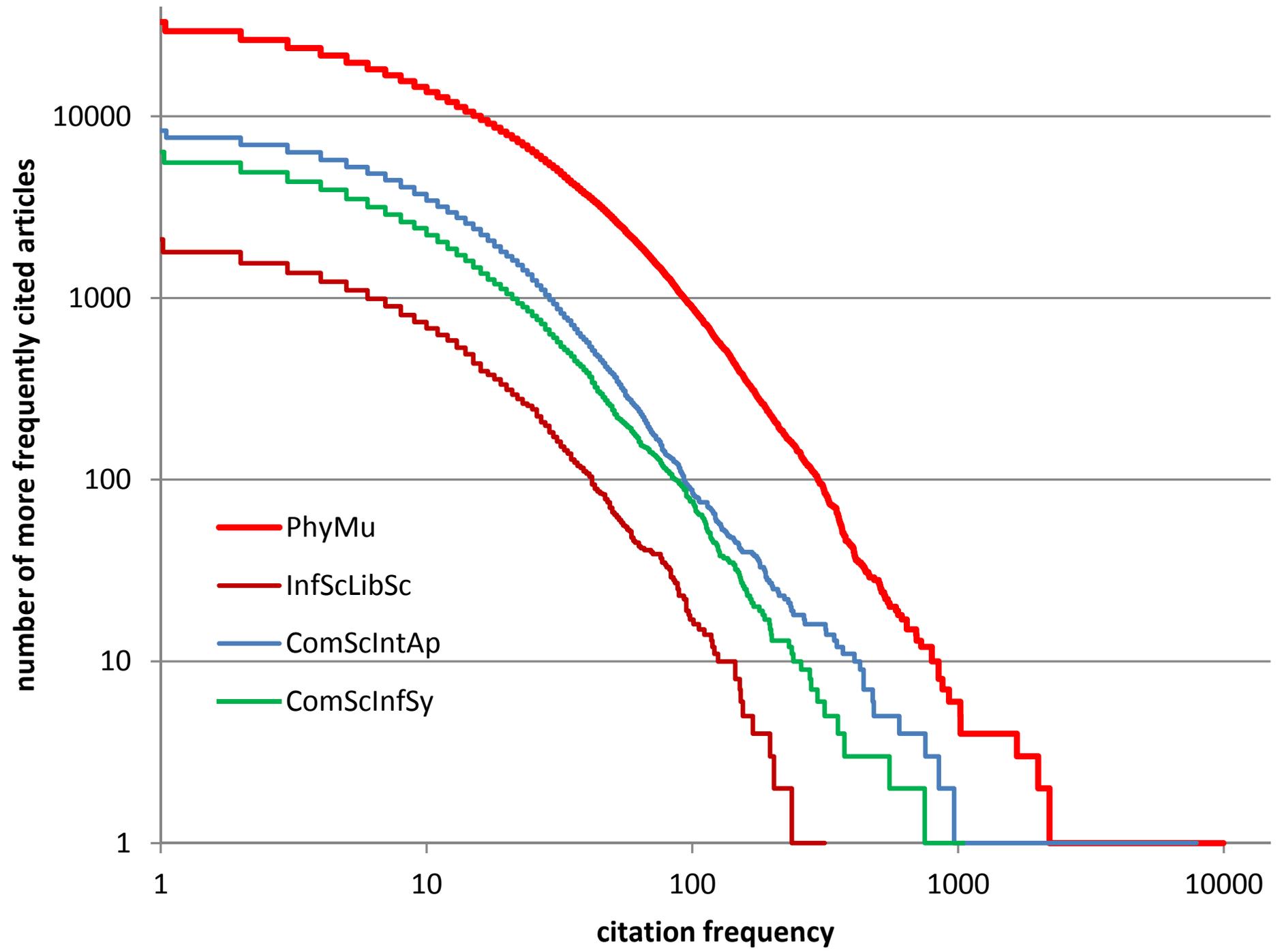

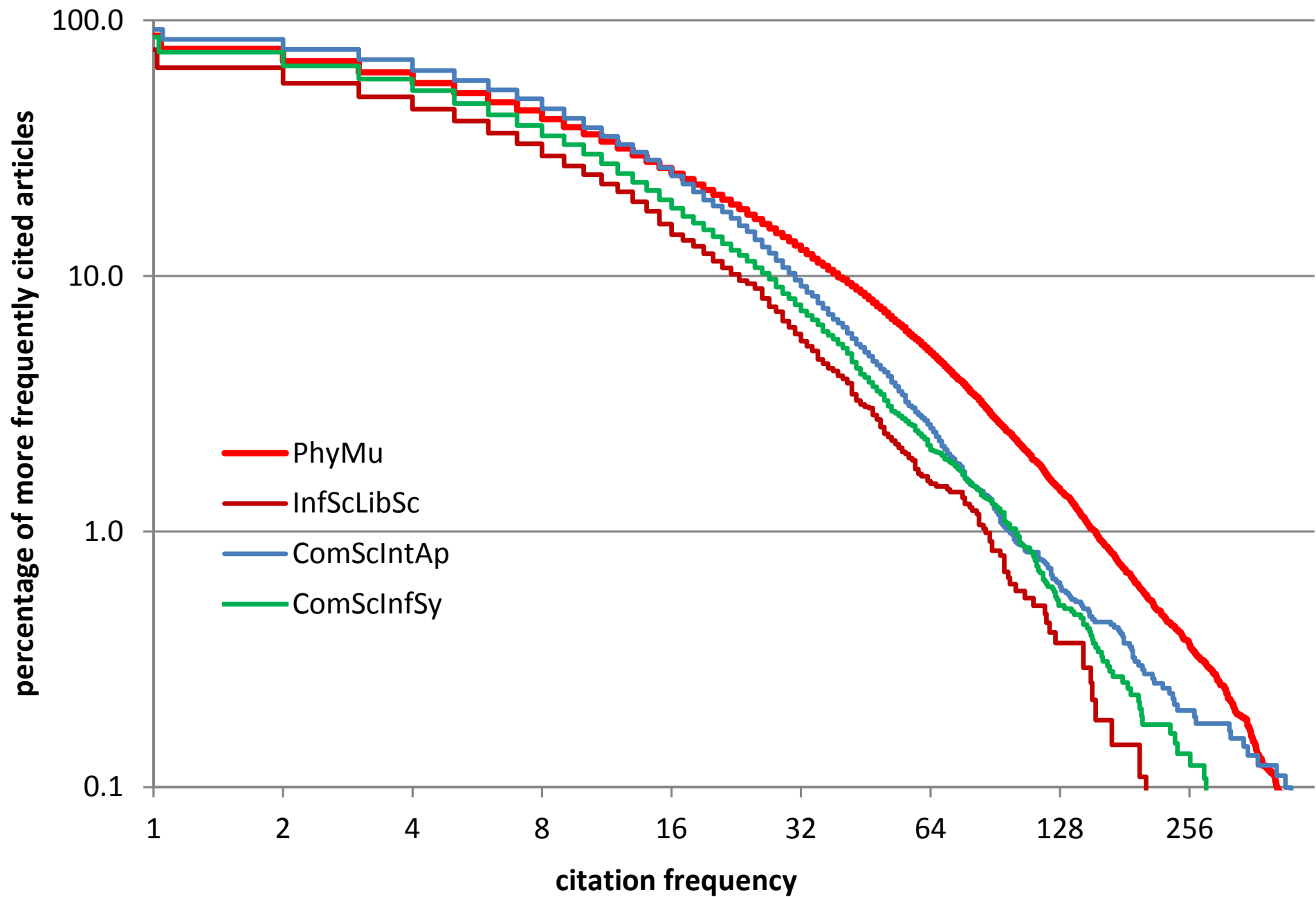

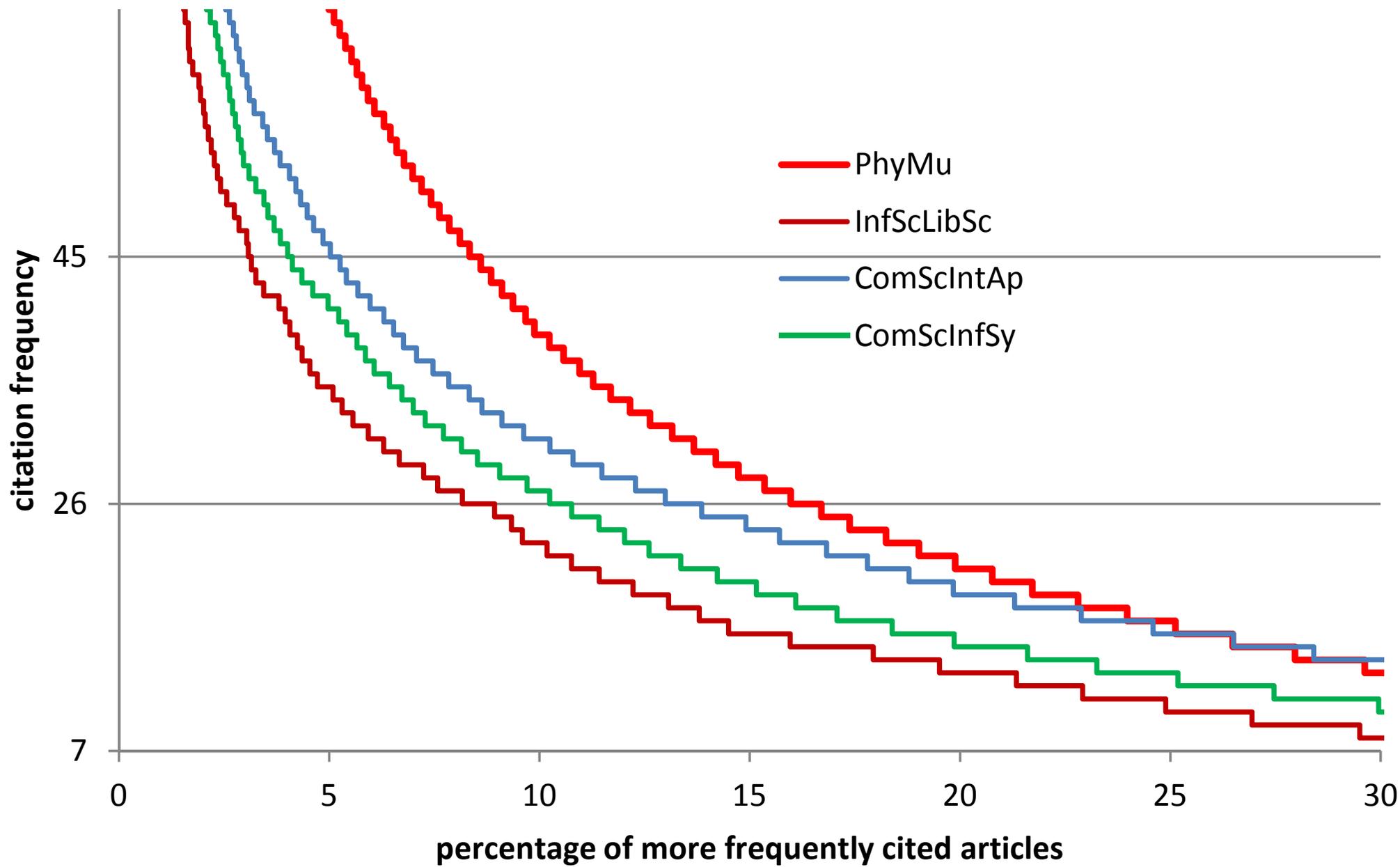

**Table 1.** Publication year, number of citations, range of ranks for the tied articles with this number of citations, total number of articles in the reference set, percentage range corresponding to that citation frequency, and median percentage for my 5 articles about variants of the Hirsch index published in multidisciplinary physics (PhyMu) journals.

| Year | No.cits. | Range of ranks | No.arts. | Percentage range | Median |
|---|---|---|---|---|---|
| 2007 | 26 | 6058 - 6333 | 37918 | 15.98 - 16.70 | 16.3 |
| 2007 | 45 | 3163 - 3261 | 37918 | 8.34 - 8.60 | 8.5 |
| 2008 | 31 | 4627 - 4812 | 41679 | 11.10 - 11.55 | 11.3 |
| 2009 | 13 | 9725 - 10424 | 40367 | 24.09 - 25.82 | 25.0 |
| 2010 | 18 | 5938 - 6321 | 39865 | 14.90 - 15.86 | 15.4 |

**Table 2.** Same as Table 1, but for the subject category Information Science / Library Science (InfScLibSc).

| Year | No.cits. | Range of ranks | No.arts. | Percentage range | Median |
|---|---|---|---|---|---|
| 2007 | 26 | 224 - 244 | 2732 | 8.20 - 8.93 | 8.6 |
| 2007 | 45 | 85 - 86 | 2732 | 3.11 - 3.15 | 3.1 |
| 2008 | 31 | 141 - 150 | 2869 | 4.91 - 5.23 | 5.1 |
| 2009 | 13 | 408 - 448 | 3079 | 13.25 - 14.55 | 13.9 |
| 2010 | 18 | 167 - 184 | 3225 | 5.18 - 5.71 | 5.4 |

**Table 3.** Same as Table 1, but for the subject category Computer Science Information Systems (ComScInfSy).

| Year | No.cits. | Range of ranks | No.arts. | Percentage range | Median |
|---|---|---|---|---|---|
| 2007 | 26 | 759 - 797 | 7402 | 10.25 - 10.77 | 10.5 |
| 2007 | 45 | 298 - 305 | 7402 | 4.03 - 4.12 | 4.1 |
| 2008 | 31 | 421 - 437 | 8040 | 5.24 - 5.44 | 5.3 |
| 2009 | 13 | 1210 - 1332 | 8458 | 14.31 - 15.75 | 15.0 |
| 2010 | 18 | 523 - 562 | 8972 | 5.83 - 6.26 | 6.0 |

**Table 4.** Same as Table 1, but for the subject category Computer Science Interdisciplinary Applications (ComScIntAp).

| Year | No.cits. | Range of ranks | No.arts. | Percentage range | Median |
|---|---|---|---|---|---|
| 2007 | 26 | 1174 - 1251 | 9029 | 13.00 - 13.86 | 13.4 |
| 2007 | 45 | 455 - 475 | 9029 | 5.04 - 5.26 | 5.2 |
| 2008 | 31 | 726 - 761 | 10061 | 7.22 - 7.56 | 7.4 |
| 2009 | 13 | 2211 - 2447 | 10825 | 20.42 - 22.61 | 21.5 |
| 2010 | 18 | 1007 - 1108 | 11006 | 9.15 - 10.07 | 9.6 |

**Table 5.** Range of ranks for the tied articles, total number of articles in the reference set, corresponding range of percentages, and median percentage for articles with 32 citations published in 2008 in the 4 subject categories from Tables 1 – 4.

| Subject category | Range of ranks | No.arts. | Percentage range | Median |
|---|---|---|---|---|
| PhyMu | 4470 - 4626 | 41679 | 10.72 - 11.10 | 10.9 |
| InfScLibSc | 127 - 140 | 2869 | 4.43 - 4.88 | 4.7 |
| ComScInfSy | 387 - 420 | 8040 | 4.81 - 5.22 | 5.0 |
| ComScIntAp | 680 - 725 | 10061 | 6.76 - 7.21 | 7.0 |